\pdfoutput=1 

\documentclass[
     a4paper,
     ]{article}

\usepackage{IEK10} 
\usepackage{natbib}

\newcommand{\mytitle}{Thermodynamics-Consistent Graph Neural Networks}

\newcommand{\affil}{
  \begin{itemize}[leftmargin=3mm, itemsep=0mm]
        \item[$^a$]RWTH Aachen University, Process Systems Engineering (AVT.SVT), Aachen, Germany %
		\item[$^b$]Forschungszentrum J\"ulich GmbH, Institute of Climate and Energy Systems ICE-1: Energy Systems Engineering, J\"ulich, Germany%
		\item[$^c$]JARA-ENERGY, Aachen, Germany
  \end{itemize}
}

\def\firstAuthor{Jan Rittig}
\newcommand{\myauthor}{
	Jan G. Rittig$^a$ \&
	Alexander Mitsos$^{b,a,c,*}$ %
}
\newcommand{\correspondingauthor}{
    Corresponding author, E-mail: amitsos@alum.mit.edu
}
\author{\myauthor}

\usepackage[hyphens]{url}
\usepackage[
  colorlinks,
  linkcolor=blue,
  citecolor=blue,
  urlcolor=blue,
  pdftitle={\mytitle},
  pdfauthor={\firstAuthor}
]{hyperref}
\usepackage[capitalise, nameinlink]{cleveref}
\crefname{table}{Tab.}{Tab.}

\begin{document}

	\thispagestyle{firststyle}
	
	\begin{center}
		\begin{large}
			\textbf{\mytitle}
		\end{large} \\
		\vspace{0.1cm}
		\myauthor
	\end{center}
	
	\vspace{-0.4cm}
	
	\begin{footnotesize}
		\affil
	\end{footnotesize}
	
	\vspace{-0.3cm}

	\section*{Abstract}
	
	We propose excess Gibbs free energy graph neural networks (GE-GNNs) for predicting composition-dependent activity coefficients of binary mixtures.
	The GE-GNN architecture ensures thermodynamic consistency by predicting the molar excess Gibbs free energy and using thermodynamic relations to obtain activity coefficients. 
	As these are differential, automatic differentiation is applied to learn the activity coefficients in an end-to-end manner.
	Since the architecture is based on fundamental thermodynamics, we do not require additional loss terms to learn thermodynamic consistency.
	As the output is a fundamental property, we neither impose thermodynamic modeling limitations and assumptions.
	We demonstrate high accuracy and thermodynamic consistency of the activity coefficient predictions.

	\vspace{0.1cm}



\section{Introduction}\label{sec:Intro}

\noindent Machine learning (ML) has shown great potential for predicting activity coefficients of binary mixtures which are highly relevant for modeling the nonideal behavior of molecules in mixtures, e.g., in separation processes. 
Various ML models such as transformers~\citep{Winter.2022}, graph neural networks (GNNs)~\citep{Felton.2022, SanchezMedina.2022, Qin.2023, Rittig.2023, SanchezMedina.2023, Zenn.2024}, and matrix completion methods (MCMs)~\citep{Chen.2021, Jirasek.2021} have been used to predict activity coefficients, exploring different representations of mixtures as strings, graphs, or matrices.
These ML models have reached high prediction accuracy beyond well-established thermodynamic models, cf.~\cite{Chen.2021, Jirasek.2021, SanchezMedina.2022, Winter.2022}, but typically lack thermodynamic consistency.

To include thermodynamic insights, ML has been combined with thermodynamic models in a hybrid fashion, e.g., in~\cite{Jirasek.2023, DiCaprio.2023, Abranches.2023, Winter.2023_PCSAFT, Felton.2024}.
Hybrid ML models promise higher predictive quality and model interpretability with less required training data.
For activity coefficients, ML has been joined with thermodynamics models such as NRTL~\citep{Renon.1968} and UNIFAC~\citep{Fredenslund.1975}, cf.~\cite{SanchezMedina.2022, Winter.2023, Jirasek.2021}.
Since thermodynamic models are associated with theoretical assumptions and corresponding limitations, the resulting hybrid models, however, also exhibit predictive limitations.

We thus recently proposed a physics-informed approach by using thermodynamic consistency equations in model training~\citep{Rittig2023_GDI}.
Physics-informed ML uses algebraic and differential relations to the prediction targets in the model architecture and training, and has already been utilized in molecular and materials property prediction, cf.~\cite{Masi.2021, Rosenberger.2022, Chaparro.2023, Chaparro.2024}.
Specifically for activity coefficients, we added the differential relationship with respect to the composition of the Gibbs-Duhem equation to the loss function of neural network training -- in addition to the prediction loss. 
Due to the high similarities to physics-informed neural networks~\citep{Raissi.2019, Karniadakis.2021}, we referred to this type of models as Gibbs-Duhem-informed neural networks.
The Gibbs-Duhem-informed GNNs and MCMs achieved high prediction accuracy and significantly increased the Gibbs-Duhem consistency of the predictions, compared to models trained on the prediction loss only.
However, this approach learns thermodynamic consistency in the form of a regularization term (also referred to as soft constraint) during training. 
It therefore requires tuning an additional parameter, i.e., weighting factor for the regularization, and does not ensure consistency.

Herein, we propose to instead use thermodynamic differential relationships \emph{directly in the activity coefficient} prediction step.
That is, the output of the ML model is the excess Gibbs free energy, a fundamental thermodynamic property.
We then utilize its relationship to the activity coefficients in binary mixtures for making predictions, thereby imposing thermodynamic consistency.
Using differential relations to the Gibbs or Helmholtz free energy has already been used in previous works to develop equation of states with ANNs.
For example, \cite{Rosenberger.2022} and \cite{Chaparro.2023} trained ANNs to predict the Helmholtz free energy with first- and second-order derivatives related to thermophysical properties, such as intensive entropies and heat capacities, by applying automatic differentiation.
They could thereby provide thermodynamics-consistent property predictions.
However, so far only properties of Lennard-Jones fluids and Mie particles have been considered by using corresponding descriptors, e.g., well depth and attractive/repulsive potentials, as input parameters to an ANN~\citep{Rosenberger.2022, Chaparro.2023, Chaparro.2024}.
To cover a diverse set of molecules, we propose to combine thermodynamic differential relations with GNNs.
We also extend previous approaches to mixture properties.
As a prime example, we combine differential relations of the excess Gibbs free energy with GNNs to predict activity coefficients of a wide spectrum of binary mixtures.
We call our models excess Gibbs free energy (GE)-GNNs.

\section{Methods \& Modeling}\label{sec:Methods}

\begin{figure}[htb]
	\centering
	\includegraphics[width=\textwidth, trim={0cm 9cm 0cm 0cm},clip]{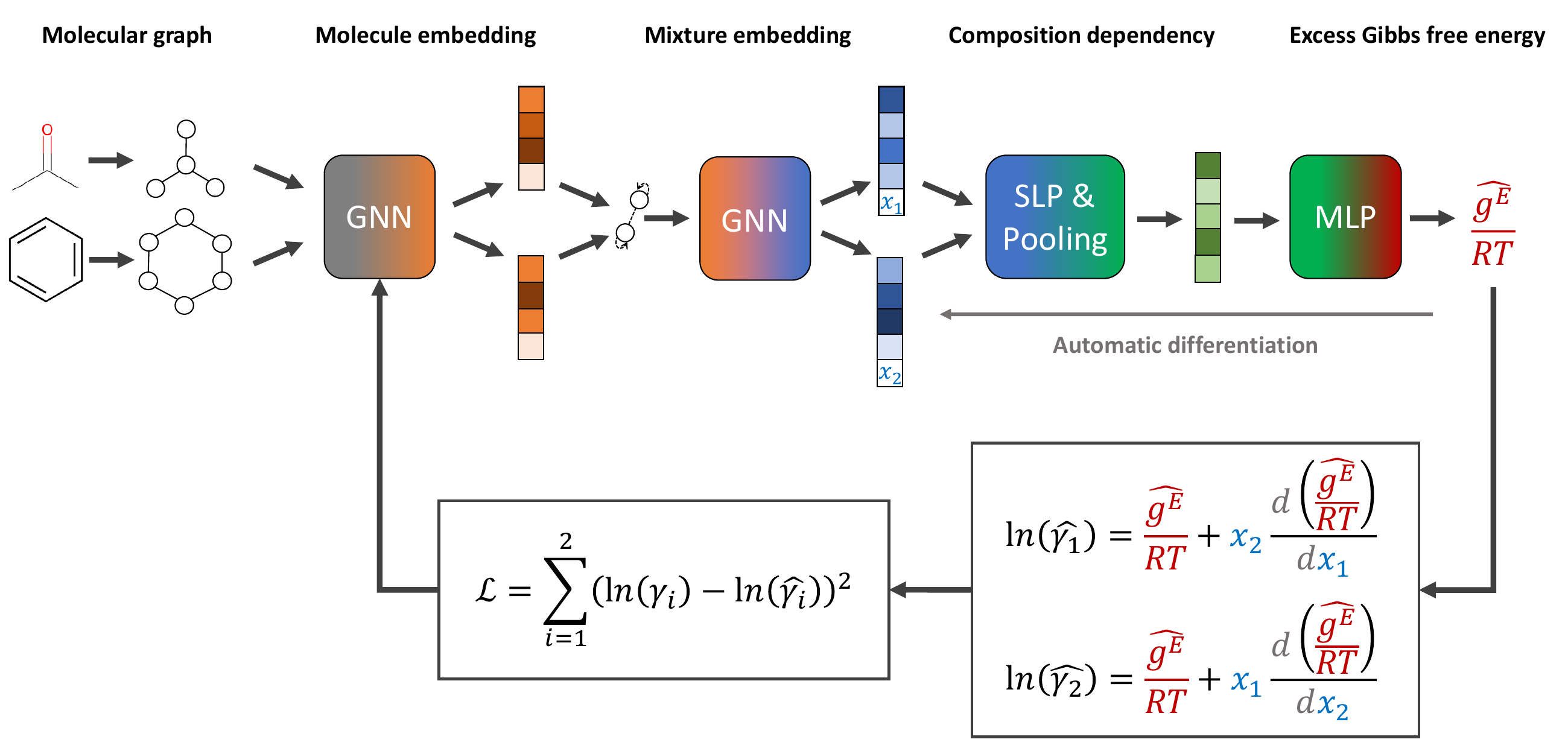}
	\caption{Model structure and loss function of our excess Gibbs free energy graph neural network (GE-GNN) for predicting composition-dependent activity coefficients.}
	\label{fig:GEGNN_structure}
\end{figure}

\noindent
The general architecture of our GE-GNNs is illustrated in Figure~\ref{fig:GEGNN_structure}.
The architecture is inspired by the SolvGNN model proposed by \cite{Qin.2023}, which we also used for our Gibbs-Duhem-informed GNNs~\citep{Rittig2023_GDI}.

\subsection{Excess Gibbs Free Energy Graph Neural Networks}
The GE-GNN takes molecular graphs as input and first learns molecular vector representations, i.e., molecular fingerprints, in graph convolutions and a pooling step; for details see overviews in~\citep{Gilmer.2017, Coley.2017, Reiser.2022, Rittig.2022_GNNBook, Schweidtmann.2023, heid2023chemprop}.
Then, a mixture graph is constructed with the components being nodes (here two nodes) that have the molecular fingerprints as node feature vectors~\citep{Qin.2023, SanchezMedina.2023, Rittig2023_GDI}.
An additional graph convolutional layer is applied on the mixture graph to capture molecular interactions, resulting in updated molecular fingerprints.
We concatenate the compositions to these fingerprints and apply single layer perceptron (SLP) with a subsequent pooling step, yielding a vector representation of the mixture, referred to as mixture fingerprint.
Lastly, an MLP takes the mixture fingerprint as input and predicts the molar excess Gibbs free energy. 

To obtain activity coefficient predictions, we utilize differential thermodynamic relationships.
Specifically, we use the relationship of the activity coefficient in binary mixtures to the molar excess Gibbs free energy (for details see Appendix):
\begin{subequations}\label{eqn:GE_gamma}
    \begin{align}
    \ln(\gamma_1) = \frac{g^E}{RT} + x_2 \frac{d (g^E / RT)}{d x_1} \label{eq:GE_gamma1} \\
    \ln(\gamma_2) = \frac{g^E}{RT} + x_1 \frac{d (g^E / RT)}{d x_2} \label{eq:GE_gamma2}
    \end{align}
\end{subequations}
Given Equ.~\ref{eq:GE_gamma1} \& \ref{eq:GE_gamma2}, we use $(g^E / RT)$ as the prediction target, corresponding to the output node of the GNN, from which we then calculate the binary activity coefficients.
The first term of the equations corresponds to the output node, while the second part, i.e., the differential term, can be calculated by using automatic differentiation of the GNN with respect to the compositions.
Then, the deviations between the predictions and the (experimental/simulated) activity coefficient data are used in the loss function.
Note that $R$ and $T$ are part of the prediction and not additional inputs, as we do not consider the temperature dependence here, which would be highly interesting for future work.
As the Gibbs free energy is a fundamental property, the derived Equ.~\ref{eq:GE_gamma1}~\&~\ref{eq:GE_gamma2} for the activity coefficients are thermodynamically consistent. 
It is trivial to check that they satisfy for instance the Gibbs-Duhem equation.

To obtain a continuously differentiable prediction curve of the activity coefficient over the composition, which is necessary for thermodynamic consistency, we apply the smooth activation function softplus for the SLP and the MLP.
We use softplus as it has been shown to be effective for molecular modeling by \cite{Schutt.2020} and in our previous work~\citep{Rittig2023_GDI}.
In fact, we found that using ReLU in the SLP/MLP can cause the model to stop learning in early epochs, resulting in very inaccurate predictions, which is presumably due to the non-smoothness of ReLU.
For more details on the effect of the activation function, we refer the interested reader to our previous work~\citep{Rittig2023_GDI}.

\subsection{Mixture Permutation Invariance}
To ensure permutation invariance with respect to the molecular inputs, we express all equations in terms of $x_1$ (i.e., $x_2 = 1 - x_1$ and $d x_1 = - d x_2$) and apply a pooling step, in contrast to simply concatenating the two molecular fingerprints, for obtaining the mixture fingerprint.
Changing the input order, e.g., ethanol/water vs. water/ethanol, thus results in the same activity coefficient predictions for the respective components.
We note that the compositions could also be concatenated to the molecular fingerprints before entering the mixture GNN model for modeling molecular interactions, without using an additional SLP to capture the composition dependency.
This requires using smooth activation functions (e.g., softplus) in the GNN part to obtain a continuously differentiable activity coefficient curve (cf.~\cite{Rittig2023_GDI}).
However, we found this alternative architecture to result in lower prediction performance.

\subsection{Training and Evaluation}
For training and evaluation, we use the composition-dependent activity coefficient data generated with COSMO-RS~\citep{Klamt.1995, Klamt.2010} by \cite{Qin.2023}.
The data set contains 280,000 activity coefficients that correspond to 40,000 binary mixtures based on the combination of 700 different compounds at seven different compositions, specifically $\{0, 0.1, 0.3, 0.5, 0.7, 0.9, 1\}$, with 0 and 1 denoting infinite dilution. 
Analogously to our previous work~\citep{Rittig2023_GDI}, we use different data split types:
\begin{itemize}
    \item[] In the \emph{comp-inter} split, activity coefficients at random compositions are excluded for some but not all mixtures, thus testing whether the model learns the composition-dependency of the activity coefficients.
    \item[] For the \emph{comp-extra} split, we exclude activity coefficients at specific compositions for all binary mixtures from training and use those for testing, e.g., $\{0.1, 0.9\}$.
    This allows us to assess the generalization capabilities to unseen compositions.
    \item[] In the \emph{mixt-extra} split, some binary mixtures are completely excluded from training and the corresponding molecules only occur in other combinations. 
    The excluded mixtures are then used for testing, thereby allowing to evaluate the generalization capabilities to new combinations of molecules.
\end{itemize}
For comp-inter and mixt-extra, we use a 5-fold stratified split based on polarity features, analogously to previous works~\citep{Qin.2023, Rittig2023_GDI}, whereas for comp-extra all compositions are excluded from training in the respective split.
The respective test sets are then used to assess the prediction quality and thermodynamic consistency.

For the predictive quality, we use the root mean squared error (RMSE), the mean absolute error (MAE), and coefficient of determination (R$^2$) of the predictions and the data.
For the thermodynamic consistency, we consider the deviation from the Gibbs-Duhem (GD) differential equation (cf. Appendix) in the form of the RMSE, i.e., referred to as GD-RMSE~\citep{Rittig2023_GDI}.
The GD-RMSE is evaluated at the compositions of the test data set, i.e., GD-RMSE$_\text{test}$, and at external compositions for which activity coefficient data is not readily available and is thus not used in training, referred to as GD-RMSE$_\text{test}^\text{ext}$.
In figures, we further consider the MAE for the Gibbs-Duhem differential equation and the molar excess Gibbs free energy.

We provide the code for the model and data splitting as open-source at~\cite{GDINN_GIT}.
To ensure comparability to previous models, we use the same model and training hyperparameters as in our previous work~\citep{Rittig2023_GDI}.

\section{Results \& Discussion}\label{sec:Results_Discussion}
\noindent Table~\ref{tab:Comp_inter} shows the prediction accuracy and Gibbs-Duhem consistency for different ML models evaluated on the comp-inter and mixt-extra splits.
The SolvGNN by \cite{Qin.2023} directly predicts activity coefficients; the model is trained on the prediction loss only, i.e., the deviation between predictions and activity coefficient data, without using thermodynamic relations.
The GDI-GNN, GDI-GNN$_\text{xMLP}$, and GDI-MCM models are different ML models from our previous work~\citep{Rittig2023_GDI} that also directly predict the activity coefficients and use the Gibbs-Duhem equation as a regularization term in the loss function during training, thereby learning but not imposing thermodynamic consistency.
The GDI model training is additionally enhanced by using a data augmentation strategy, that is, the deviation from the Gibbs-Duhem differential relationships at random compositions (not only at the compositions for which activity coefficients are available for training) are also considered in training, so that the models can learn thermodynamic consistency over the whole composition range.
We compare these models to the GE-GNN proposed in this work.

\begin{table}[b!]
    \centering
    \caption{Comparison of prediction accuracy and Gibbs-Duhem consistency for comp-inter and mixt-extra data split using different machine learning models. Bold print indicates best performance.}
	\resizebox{\linewidth}{!}{%
    \begin{tabular}{l|ccc|ccc}
        \toprule
        \multicolumn{1}{c|}{\multirow{2}[3]{*}{\shortstack[c]{Model}}}& \multicolumn{3}{c|}{comp-inter} & \multicolumn{3}{c}{mixt-extra} \\
         & RMSE$_\text{test}$ &  GD-RMSE$_\text{test}$ & GD-RMSE$_\text{test}^\text{ext}$  & RMSE$_\text{test}$ &  GD-RMSE$_\text{test}$ & GD-RMSE$_\text{test}^\text{ext}$ \\
         \midrule 
         SolvGNN~\citep{Qin.2023}$^\dagger$ & 0.088 & 0.212 & 0.298 & 0.114 & 0.206 & 0.311 \\
         GDI-GNN~\citep{Rittig2023_GDI} & 0.081 & 0.032 & 0.038 & \textbf{0.105} & 0.040 & 0.038 \\
         GDI-GNN$_\text{xMLP}$~\citep{Rittig2023_GDI} & 0.083 & 0.028 & 0.025 & 0.113 & 0.035 & 0.030 \\
         GDI-MCM~\citep{Rittig2023_GDI}  & 0.088 & 0.034 & 0.035  & 0.120 & 0.039 & 0.036 \\
         \hline
         GE-GNN (this work) & \textbf{0.068} & \textbf{0.000} & \textbf{0.000}  & 0.114 & \textbf{0.000} & \textbf{0.000} \\
        \bottomrule[.5pt]
        \multicolumn{4}{l}{$^{\dagger}$\footnotesize{Model was reevaluated in~\citep{Rittig2023_GDI}.}}\\
    \end{tabular}}
    \label{tab:Comp_inter}
\end{table}

The results show that the GE-GNN model outperforms the other models by achieving a higher prediction accuracy of 0.068 RMSE on the comp-inter test set.
The GE-GNN further imposes Gibbs-Duhem consistency, i.e., exhibits a GD-RMSE$_\text{test}$ and a GD-RMSE$_\text{test}^\text{ext}$ of 0.
For the mixt-extra sets, the GDI-GNN shows the highest prediction accuracy with an RMSE of 0.105, whereas the GE-GNN exhibits a slightly worse RMSE of 0.114, but indeed preservers thermodynamic consistency.

\begin{figure}
    \includegraphics[width=\textwidth, height=0.85\textheight, trim={0cm 10cm 0cm 0cm},clip]{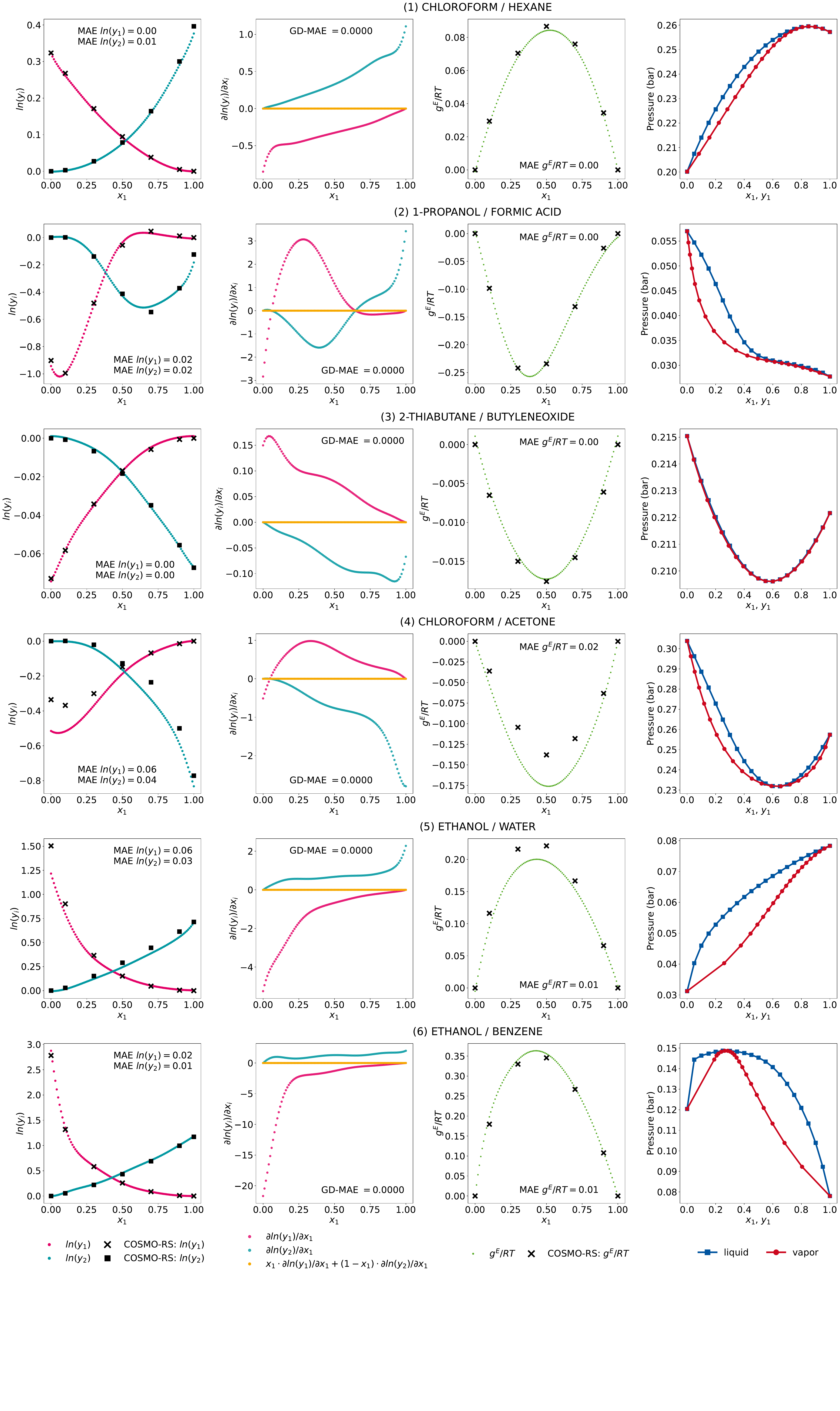}
    \caption{Activity coefficient predictions, their corresponding gradients with respect to the composition with the associated Gibbs-Duhem deviations, the molar excess Gibbs free energy, and vapor-liquid equilibria for exemplary mixtures by the GE-GNN. The predictions are averaged from the five model runs of the comp-inter split, i.e., an ensemble.}
    \label{fig:GE_GNN_system_inter_examples}
\end{figure} 

We further show the GE-GNN's activity coefficient predictions, the corresponding gradients with respect to the composition, the molar excess Gibbs free energy, and the vapor-liquid-equilibrium (VLE) plots at 298\,K for some exemplary mixtures in Figure~\ref{fig:GE_GNN_system_inter_examples}.
We took the same exemplary mixtures as in our previous work on GDI-GNNs (cf.~\cite{Rittig2023_GDI}) to ensure comparability and reflect different nonideal behaviors in binary mixtures, hence different activity coefficient curves. 
The VLEs are obtained using Raoult's law and the Antoine equation with parameters from the National Institute of Standards and Technology (NIST) Chemistry webbook~\citep{Linstrom.2001} based on the work by \cite{Qin.2023} and \cite{NIST_scrap}.

We observe accurate predictions of the activity coefficients that are consistent with the Gibbs-Duhem equation for all mixtures.
In particular, for systems (1)-(3) and (6), the predicted activity coefficients match the COSMO-RS data very accurately, which is also reflected in an accurate fit of the molar excess Gibbs free energy.
For systems (4) and (5), i.e., chloroform/acetone and ethanol/water, the infinite dilution activity coefficients for the second component ($x_1 \rightarrow 1$) show some deviations.
For these systems, we also find slight deviations in the activity coefficient predictions at intermediate compositions, which leads to an underestimation of the molar excess Gibbs free energies in both cases.
Yet, the general trend in the activity coefficient and corresponding molar excess Gibbs free energies curves is well captured.
Furthermore, we observe thermodynamically consistent and smooth VLE plots for all systems, which we have shown to be problematic when ML models are trained only on activity coefficients without using thermodynamic insights, cf.~\cite{Rittig2023_GDI}.
The GE-GNNs are therefore able to capture various nonideal behaviors in the exemplary mixtures with thermodynamic consistency and provide overall highly accurate predictions.

In addition, we report the prediction accuracy and thermodynamic consistency for the comp-extra set in Table~\ref{tab:comp-extra}, where we exclude specific compositions for all mixtures from the training set and use them for testing (cf. Section~\ref{sec:Methods}).
We note that this scenario is rather artificial and aims to test the generalization capabilities in an extreme case. 
In practice, experimental data for these compositions is readily available.
We compare the GE-GNN with the same models as for the comp-inter and mixt-extra split.

\begin{table}[!b]
	\caption{Comparison of prediction accuracy and Gibbs-Duhem consistency for comp-extra split, i.e.,  specific compositions excluded from training and used for testing (first row), using different machine learning models. Bold print indicates best performance.}
	\label{tab:comp-extra}
	\resizebox{\linewidth}{!}{%
		\begin{tabular}{l|rr|rr|rr|rr}
			\toprule
			\multicolumn{1}{c|}{\multirow{2}[3]{*}{\shortstack[c]{Model}}} & \multicolumn{2}{c|}{excl. $x_i \in \{0.5\}$} & \multicolumn{2}{c|}{excl. $x_i\in \{0.3, 0.7\}$} & \multicolumn{2}{c|}{excl. $x_i \in \{0.1, 0.9\}$} & \multicolumn{2}{c}{excl. $x_i \in \{0, 1\}$} \\
			 & RMSE$_\text{test}$ & GD-RMSE$_\text{test}$ & RMSE$_\text{test}$ & GD-RMSE$_\text{test}$ & RMSE$_\text{test}$ & GD-RMSE$_\text{test}$ & RMSE$_\text{test}$ & GD-RMSE$_\text{test}$ \\
			\midrule
			
			SolvGNN~\citep{Qin.2023} &  0.067 &  0.453 &  0.180 &  1.532 &  0.302 &  0.715 &  0.514 &  0.101 \\
			GDI-GNN~\citep{Rittig2023_GDI} &  0.040 &  0.030 &  0.064 &  0.034 &  \textbf{0.075} &  0.044 &  0.374 &  0.026 \\
			GDI-GNN$_\text{xMLP}$~\citep{Rittig2023_GDI} & 0.039 &  0.021 &  0.065 &  0.028 &  0.087 &  0.032 &  \textbf{0.332} &  0.044 \\
			GDI-MCM~\citep{Rittig2023_GDI} &  0.043 &  0.039 &  0.067 &  0.042 &  0.094 &  0.036 &  0.342 &  0.051 \\
			\midrule
			GE-GNN (this work)  &  \textbf{0.026} &  \textbf{0.000} &  \textbf{0.054} &  \textbf{0.000} &  0.085 &  \textbf{0.000} &  0.504 &  \textbf{0.000} \\
            \bottomrule[.5pt]
            \multicolumn{4}{l}{$^{\dagger}$\footnotesize{Model was reevaluated in~\citep{Rittig2023_GDI}.}}\\
	\end{tabular}}
\end{table}

We observe again that the GE-GNN, being thermodynamically consistent, outperforms the other models in terms of the GD-RMSE$_\text{test}$.
For the accuracy of the predictions, RMSE$_\text{test}$, we see competitive performance of the GE-GNN for intermediate compositions. 
For $x_i = 0.5$ and $x_i \in \{0.3, 0.7\}$, the GE-GNN shows superior accuracy; for $x_i \in \{0.1, 0.9\}$, the GDI-GNN performs slightly better.
In the case of infinite dilution activity coefficients ($x_i \in \{0, 1\}$), the GE-GNN is outperformed by the GDI models.

\begin{figure}
	\begin{subfigure}[c]{\textwidth}
		\centering
		\includegraphics[width=\textwidth, trim={0cm 13cm 0cm 0cm},clip]{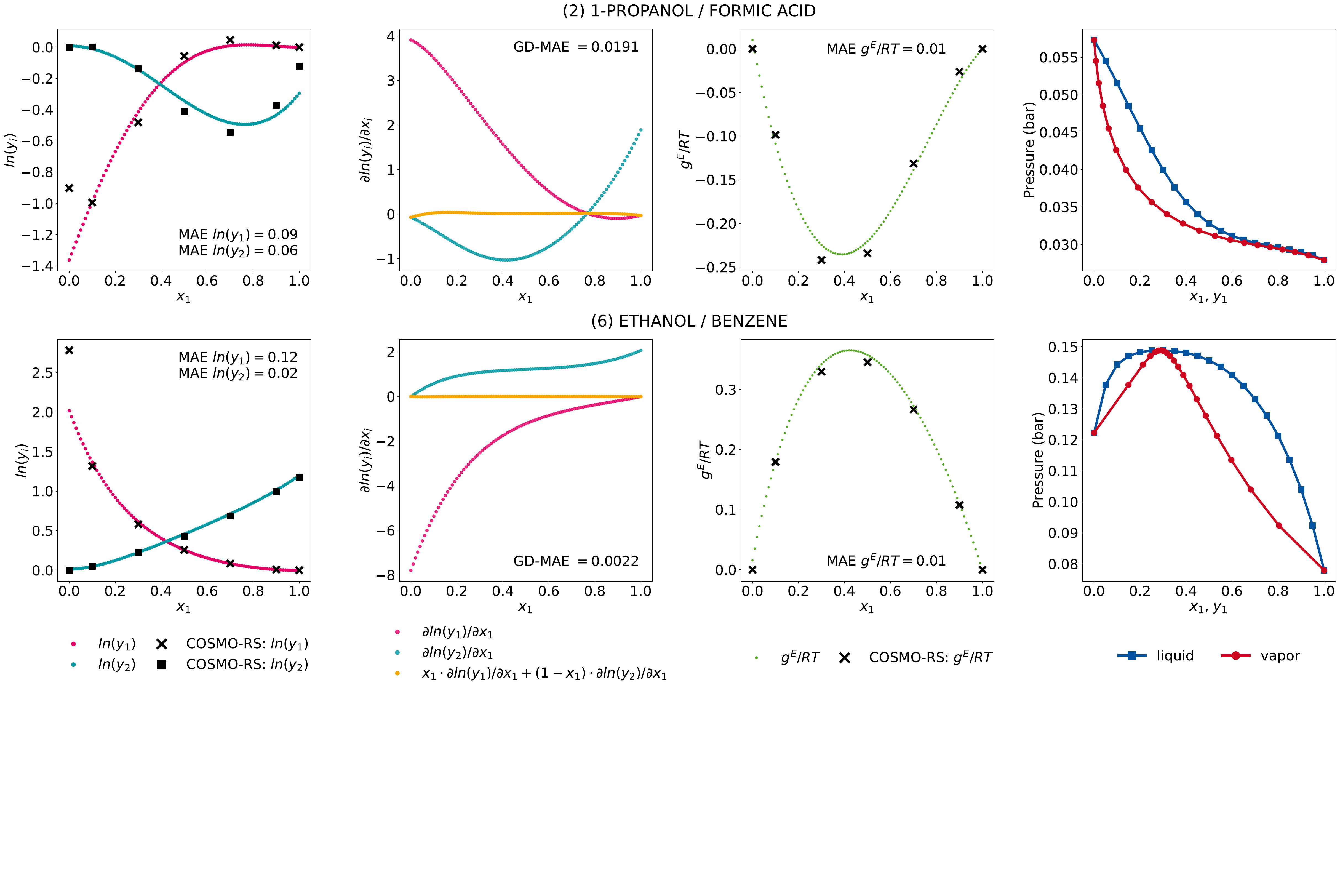}
		\subcaption{GDI-GNN$_\text{xMLP}$}
        \vspace{0.4cm}
        \label{subfig:comp_extra_GDIGNN}
	\end{subfigure}
	\begin{subfigure}[c]{1\textwidth}
		\centering
    \includegraphics[width=\textwidth, trim={0cm 13cm 0cm 0cm},clip]{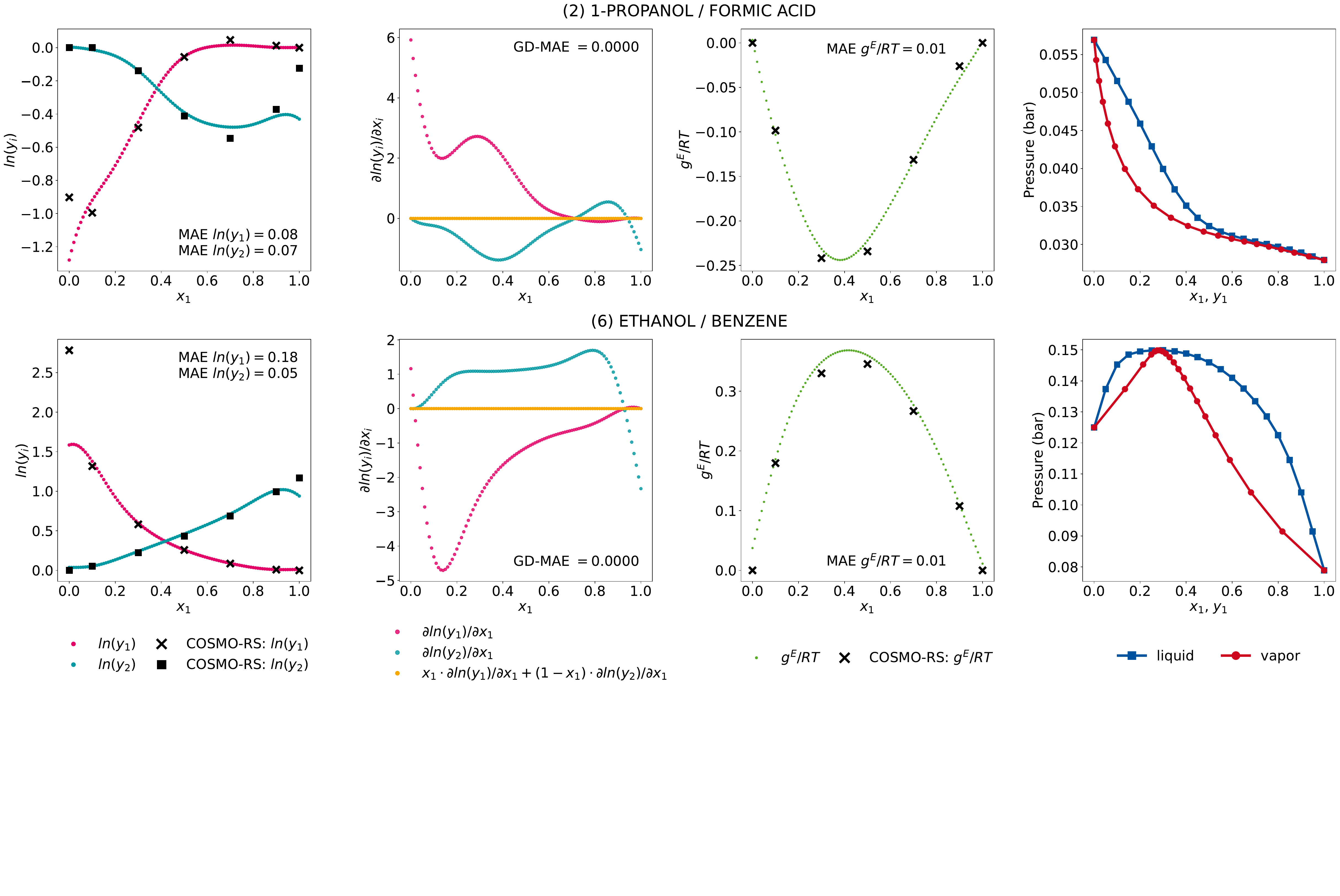}
		\subcaption{GE-GNN}
		\label{subfig:comp_extra_GEGNN}
	\end{subfigure}
	\caption{Activity coefficient predictions, their corresponding gradients with respect to the composition with the associated Gibbs-Duhem deviations, the molar excess Gibbs free energy, and vapor-liquid equilibria for the exemplary mixture of ethanol/benzene by the (a) GDI-GNN$_\text{xMLP}$ and (b) GE-GNN.}
	\label{fig:comp_extra}
\end{figure} 

To further investigate the lower accuracy of the GE-GNN for infinite dilution activity coefficients, we show two examples of ethanol/benzene and 1-propanol/formic acid of the comp-extra set for both the GDI-GNN$_\text{xMLP}$ and the GE-GNN in Figure~\ref{fig:comp_extra}.
Notably, the slopes of activity coefficients curves predicted by GDI-GNN$_\text{xMLP}$ continue for $x_i \rightarrow \{0, 1\}$.
In contrast, the GE-GNN exhibits rather drastic changes in the gradients with respect to compositions in these regions, hence not continuing the slope.
We explain this by the fact that the GE-GNN is not trained for these compositions at all and thus cannot interpolate as for intermediate compositions, hence is not sensitive in these regions of extrapolation.
The GDI-GNN$_\text{xMLP}$ is trained on Gibbs-Duhem consistency for the whole composition range, i.e., $[0,1]$, and seems to learn that having less abrupt variations in the gradients is a way to promote consistency. 
For binary mixtures, where the infinite dilution activity coefficients can be approximated by a continuation of the nonideal behavior, as for ethanol/benzene, the GDI models yield more accurate predictions.
But when binary mixtures exhibit changes in the nonideal behavior for $x_i \rightarrow \{0, 1\}$, as here 1-propanol/formic acid, both approaches fail to capture these changes, which is expected since they are not trained for these compositions.
Therefore, the higher predictive accuracy of the GDI models is presumably due to the fraction of binary mixtures for which the infinite dilution activity coefficients can be approximated by the continuation of the nonideal behavior. 
As in practice infinite dilution activity coefficients would indeed be utilized for training and it is also possible to include additional data for $x_i = 1$ with $\gamma_i = 1$, i.e., $\ln(\gamma_i) = 0$, the GNNs can learn this nonideal behavior.
Here, it would rather be interesting to extend neural network architectures, including GNNs, to impose this definition of the activity coefficient at $x_i = 1$.

\section{Conclusion}\label{sec:Conclusion}

\noindent We propose to combine GNNs with thermodynamic differential relationships between properties for binary activity coefficient prediction to ensure thermodynamic consistency.
That is, our GE-GNN predicts the excess Gibbs free energy and utilizes the relationship to activity coefficients via automatic differentiation during model training, enabling end-to-end learning of activity coefficients.
By using a fundamental property as the model output, we do not impose any thermodynamic modeling limitations or assumptions, as opposed to previously proposed ML methods.
We further do not need to learn thermodynamic consistency during training, as in physics-informed neural network approaches, which require tuning weighting factors for regularization and do not ensure consistency.
Our results show that the GE-GNNs achieve high prediction accuracy and by design exhibit Gibbs-Duhem consistency.

Incorporating additional thermodynamic insights by means of constraining the neural network architecture, e.g., $\gamma_i = 1$ for $x_i = 1$, should be addressed in future work.
It would also be interesting to capture the temperature-dependency of activity coefficients, e.g., by combining the Gibbs-Helmholtz~\citep{SanchezMedina.2023} with GE-GNNs or directly using the temperature relation in the excess Gibbs free energy.
In general, utilizing further fundamental thermodynamic algebraic/differential relationships is highly promising for future work on combining ML with thermodynamics.

\section*{Acknowledgments}

\noindent This project was funded by the Deutsche Forschungsgemeinschaft (DFG, German Research Foundation) – 466417970 – within the Priority Programme ``SPP 2331: Machine Learning in Chemical Engineering''. 
This work was also performed as part of the Helmholtz School for Data Science in Life, Earth and Energy (HDS-LEE). 
Simulations were performed with computing resources granted by RWTH Aachen University under project ``rwth1232''.
We further gratefully acknowledge Victor Zavala's research group at the University of Wisconsin-Madison for making the SolvGNN implementation and the COSMO-RS activity coefficient data openly available.

\section*{Authors contributions}
\noindent J.G.R. developed the concept of excess Gibbs free energy graph neural networks, implemented them, set up and conducted the computational experiments including the formal analysis and visualization, and wrote the original draft of the manuscript.
A.M. acquired funding, provided supervision, and edited the manuscript.


\section*{Appendix}\label{sec:App}

\noindent The relationship of the molar excess Gibbs free energy and activity coefficients we utilize can be derived from:
\begin{equation}\label{eq:GE}
\frac{g^E}{RT} =  x_1 \ln(\gamma_1) + x_2 \ln(\gamma_2)
\end{equation}
Differentiating Equ.~\ref{eq:GE} with respect to $x_1$ gives
\begin{equation*}\label{eq:dGE_x1}
\frac{d (g^E / RT)}{d x_1} = x_1 \frac{\ln(\gamma_1)}{\partial x_1} + \ln(\gamma_1) + x_2 \frac{\partial \ln(\gamma_2)}{\partial x_1} + \ln(\gamma_2) \frac{\partial x_2}{\partial x_1}.
\end{equation*}
Further inserting the Gibbs-Duhem equation for binary mixtures, i.e., 
\begin{equation*}\label{eq:GD-diff}
x_1 \cdot \left(\frac{\partial \ln({\gamma_1})}{\partial x_1}\right)_{T,p} + x_2 \cdot \left(\frac{\partial \ln({\gamma_2})}{\partial x_1}\right)_{T,p} = 0
\end{equation*}
and using $d x_1 =  - d x_2$ yields
\begin{equation}\label{eq:dGE}
\frac{d (g^E / RT)}{d x_1} = \ln \frac{\gamma_1}{\gamma_2}.
\end{equation}
Combining Equ.~\ref{eq:GE} and Equ.~\ref{eq:dGE} gives expressions for the binary activity coefficients:
\begin{align*}
    \ln(\gamma_1) = \frac{g^E}{RT} + x_2 \frac{d (g^E / RT)}{d x_1} \\
    \ln(\gamma_2) = \frac{g^E}{RT} + x_1 \frac{d (g^E / RT)}{d x_2} 
\end{align*}

  \clearpage

  \bibliographystyle{apalike}
  \renewcommand{\refname}{Bibliography}
  \bibliography{literature.bib}

\end{document}